# A Short Review on Graphene-Based Filters: Perspectives and Challenges

**Mohammad Bagher Heydari** [1,*], **Mohammad Hashem Vadjed Samiei** [1]

[1] School of Electrical Engineering, Iran University of Science and Technology (IUST), Tehran, Iran

[*] Corresponding author: mo_heydari@elec.iust.ac.ir

**Abstract:** Graphene is an interesting debated topic between scientists because of its unique properties such as tunable conductivity. Graphene conductivity can be varied by either electrostatic or magnetostatic gating or via chemical doping, which leads to design of various photonic and electronic devices. Among various graphene-based structures, plasmonic graphene filters have attracted the attention of many researchers because of their various applications in THz frequencies. There are four main types of graphene filters proposed in the literature, which are: 1- Circular filters, 2- Planar filters, 3- Periodic filters, and 4- Other types of filters (which are not categorized in any of the other three groups). This paper aims to study and investigate various types of graphene-based filters published in the literature.

## 1. Introduction

Plasmonics is a new emerging science in recent years, which has many fascinating properties in photonics and electronics [1, 2]. Both metals and two-dimensional (2D) materials can support surface plasmon polaritons (SPPs). In metals, SPPs often are excited in the visible and near-infrared regions [3, 4]. The emerging 2D materials like molybdenum disulfide ($MoS_2$) and tungsten diselenide ($WSe_2$) have generated immense interest for semiconductor and nanotechnology due to their fascinating properties [5-9]. Among these 2D materials, carbon nanotubes exhibit attractive features which are widely used in chemistry and physics [10, 11]. Graphene is a 2D sheet which offers a number of fundamentally superior properties that have the potential to enable a new generation of technologies [12-14].

The graphene research has grown slowly in the late 20th century but AB initio calculations illustrated that a single graphene layer is unstable [15]. Andre Geim and Konstantin Novoselov first isolated single layer samples from graphite in 2007 and the Nobel prize in physics 2010 was awarded to them [16, 17].

Graphene is one of the allotropes of carbon, which is a planar monolayer of carbon atoms that form a honeycomb lattice with a carbon-carbon bond length of 0.142 nm [18]. It exhibits a large wide of interesting properties [19-24]. One of them is its highly unusual nature of charge carriers which behave as Dirac fermions [25]. This feature has a great effect on the energy spectrum of Landau levels produced in presence of a magnetic field [26, 27]. The Hall conductivity is observed at zero energy level. Also, the Hall Effect is distinctly different than the conventional Hall Effect which is quantized by a half-integer [28, 29]. Graphene absorbs 2.3% of incident light and the absorption can be tuned by varying the Fermi level through the electrical gating [30, 31]. This property is widely used in designing transparent electrodes in solar cells [32-34]. Graphene has an exceptional thermal conductivity (5000 w/m$^{-1}$ K$^{-1}$) that is utilized in fabricating and designing the electronic components [35].

Graphene transport characteristics and electrical conductivity can be tuned by either electrostatic or magnetostatic gating or via chemical doping, which this conductivity leads to the design of various photonic and electronic devices [2, 36-42]. All of the mentioned fascinating features make the graphene as a good candidate for designing novel devices in the THz frequencies such as waveguides [43-49], Radar Cross-Section (RCS) reduction-based devices [50-52], and graphene-based medical devices [53-59]. Furthermore, integration graphene with anisotropic materials such as ferrites, which have many interesting properties in the microwave regime [60, 61], can enhance the performance of the graphene-based devices [44]. One of the interesting graphene-based device designed by graphene conductivity is a plasmonic filter which has various applications in communications. This paper presents a short review of plasmonic graphene filters.

The paper is organized as follows. Due to the importance of graphene conductivity in designing plasmonic filters, the conductivity of graphene in presence of electrostatic or magnetostatic bias has been considered in section 2. Then, various types of graphene filters are introduced and their performance principles are discussed in section 3. These filters can be categorized into four groups: 1- Circular filters, 2- Planar filters, 3- Periodic filters, and 4- Other types of filters. Section 4 briefly considers the perspective and challenges. Finally, section 5 concludes the paper.

## 2. The Conductivity of Graphene

In graphene plasmonics, the conductivity of graphene plays the main role in designing photonic devices, because it describes the electromagnetic interactions between the graphene sheet and the external field. The popular relation used for the conductivity has been computed via Kubo's formula [62]:

$$\sigma(\omega,\mu_c,\Gamma,T) = \frac{je^2(\omega - j2\Gamma)}{\pi \hbar^2} \left[ \begin{array}{l} \frac{1}{(\omega - j2\Gamma)^2} \int_0^\infty \varepsilon \left( \frac{\partial f_d(\varepsilon)}{\partial \varepsilon} - \frac{\partial f_d(-\varepsilon)}{\partial \varepsilon} \right) d\varepsilon \\ - \int_0^\infty \frac{f_d(\varepsilon) - f_d(-\varepsilon)}{(\omega - j2\Gamma)^2 - 4(\varepsilon/\hbar)^2} \left( \frac{\partial f_d(\varepsilon)}{\partial \varepsilon} - \frac{\partial f_d(-\varepsilon)}{\partial \varepsilon} \right) d\varepsilon \end{array} \right] \quad (1)$$

Where

$$f_d(\varepsilon) = \frac{1}{1 + \exp\left(\frac{\varepsilon - \mu_c}{K_B T}\right)} \quad (2)$$

In the above equations, $e$ is the charge of an electron, $\hbar$ is the reduced Planck's constant, $K_B$ is Boltzmann's constant, $T$ is temperature, $\omega$ is radian frequency, and $\mu_c$ is chemical potential. By substituting (2) into (1) and integrating, one can obtain [62]:

$$\sigma(\omega,\mu_c,\Gamma,T) = \sigma_{inter}(\omega) + \sigma_{intra}(\omega) = \frac{-je^2}{4\pi \hbar} Ln \left[ \frac{2|\mu_c| - (\omega - j2\Gamma)\hbar}{2|\mu_c| + (\omega - j2\Gamma)\hbar} \right] +$$

$$\frac{-je^2 K_B T}{\pi \hbar^2 (\omega - j2\Gamma)} \left[ \frac{\mu_c}{K_B T} + 2Ln\left(1 + e^{-\mu_c/K_B T}\right) \right] \quad (3)$$

In (3), the conductivity has been split into two terms: intra-band and inter-band transitions. It should be noted that the intra-band electronic process becomes dominant in the mid-infrared and THz frequencies at room temperature. All of the plasmonic graphene couplers utilize the relation (3) for their design except non-reciprocal couplers. In the existence of the external magnetic field, the conductivity of graphene converts to tensor [63]:

$$\bar{\bar{\sigma}}(\omega,\mu_c,\Gamma,T,\vec{B}_0) = \begin{pmatrix} \sigma_d & -\sigma_o \\ \sigma_o & \sigma_d \end{pmatrix} \quad (4)$$

Where $\sigma_d, \sigma_o$ are [63]:

$$\sigma_d(\mu_c, B_0) = \frac{e^2 v_f^2 |eB_0|(\omega - j2\Gamma)\hbar}{-j\pi} \times$$

$$\sum_{n=0}^{\infty} \left[ \frac{f_d(M_n) - f_d(M_{n+1}) + f_d(-M_{n+1}) - f_d(-M_n)}{(M_{n+1} - M_n)^2 - (\omega - j2\Gamma)^2 \hbar^2} \times \left(1 - \frac{\Delta^2}{M_n M_{n+1}}\right) \times \frac{1}{M_{n+1} - M_n} \right.$$

$$\left. + \frac{f_d(-M_n) - f_d(M_{n+1}) + f_d(-M_{n+1}) - f_d(M_n)}{(M_{n+1} + M_n)^2 - (\omega - j2\Gamma)^2 \hbar^2} \times \left(1 + \frac{\Delta^2}{M_n M_{n+1}}\right) \times \frac{1}{M_{n+1} + M_n} \right] \quad (5)$$

$$\sigma_o(\mu_c, B_0) = \frac{e^2 v_f^2 |eB_0|}{\pi} \times$$

$$\sum_{n=0}^{\infty} \left[ \frac{f_d(M_n) - f_d(M_{n+1}) - f_d(-M_{n+1}) + f_d(-M_n)}{(M_{n+1} - M_n)^2 - (\omega - j2\Gamma)^2 \hbar^2} \times \left(1 - \frac{\Delta^2}{M_n M_{n+1}}\right) \right] \tag{6}$$

$$+ \left[ \frac{f_d(M_n) - f_d(M_{n+1}) - f_d(-M_{n+1}) + f_d(-M_n)}{(M_{n+1} + M_n)^2 - (\omega - j2\Gamma)^2 \hbar^2} \times \left(1 + \frac{\Delta^2}{M_n M_{n+1}}\right) \right]$$

In the above relations, $f_d$ has been defined in (2) and,

$$M_n = \sqrt{\Delta^2 + 2n v_f^2 |eB_0| \hbar} \tag{7}$$

In (7), $v_f \approx 10^6 \, m/s$ is the electron velocity in graphene, $B_0$ is applied magnetic field, and $\Delta$ is an excitonic energy gap. Relations (5) and (6) become simple for $\mu_c \gg \hbar \omega_c$ and $\mu_c \gg K_B T$ [64]:

$$\sigma_d(\omega, \mu_c, \tau, T, B_0) = \sigma_0 \frac{1 + j\omega\tau}{(\omega_c \tau)^2 + (1 + j\omega\tau)^2} \tag{8}$$

$$\sigma_o(\omega, \mu_c, \tau, T, B_0) = \sigma_0 \frac{\omega_c \tau}{(\omega_c \tau)^2 + (1 + j\omega\tau)^2} \tag{9}$$

Where

$$\omega_c = \frac{eB_0 v_f^2}{|\mu_c|} \tag{10}$$

is cyclotron frequency and the static conductivity for $B_0 = 0$ is expressed as [64]:

$$\sigma_0 = \frac{e^2 \mu_c \tau}{\pi \hbar^2} \tag{11}$$

In (11), $\tau$ is scattering time which is defined as [64]:

$$\tau = \frac{\pi \hbar^2 n_s \mu}{e \mu_c} \tag{12}$$

Where $\mu$ is the DC mobility of graphene and $n_s$ is the surface carrier density.

## 3. Various Types of Plasmonic Graphene Filters

In the literature, various types of graphene filters have been proposed. Here, we categorize them into four main groups, as illustrated in Fig. 1. This section tends to study each kind of graphene filter in a separate sub-section.

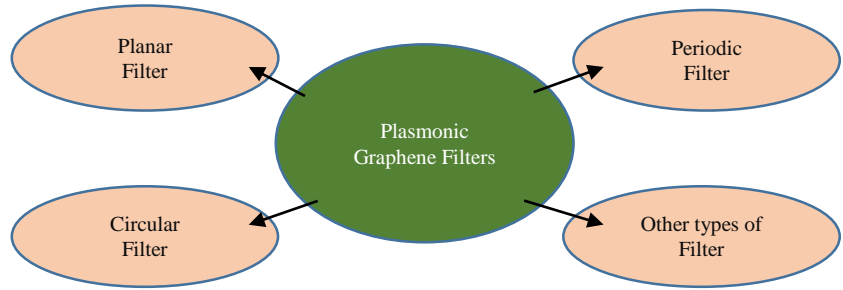

**Fig 1**. Various types of plasmonic graphene filters

*3.1 Circular Filters*

Circular filters can be divided into two groups: 1- Ring filters, 2- Disk filters. Here, we will review articles about graphene resonators, which are utilized as graphene plasmonic filters. A ring resonator is a kind of waveguide in which the input power is coupled to at least one closed-loop and gets out from the output bus. In the optics region, the optical ring resonators work based on one of the three main principles: 1- optical coupling, 2- total internal reflection, and 3- interference. The ring resonator can work as a plasmonic filter because only some wavelengths will be at the resonance of the loop.

Two research articles for ring filters based on graphene are reported in the literature. Hong-Ju li et al. proposed a band-stop filter where is composed of a graphene ribbon coupled to a ring resonator [65], as seen in fig. 2 (a). FDTD method was applied to numerically study the filter behavior in which the results revealed that edge modes could increase the coupling between ribbon and ring resonator [65]. The transmission spectrum as a function of wavelength for different outer radius of graphene rings has been shown in fig.3 (a) [65]. This figure indicates that redshifts occur for transmission dips as the outer radius increases [65]. The mentioned filter can be utilized in future integrated circuits due to its tunability [65]. A similar structure for tunable filter based on graphene split-ring resonator has been investigated by Yixiao Gao and his coworkers [66]. This filter was realized by a split on a graphene ring with sector angle $\theta$ and graphene strip. The authors used FEM to simulate and study the tunable filter which the achieved results indicated that transmission resonances could be divided into two main categories: even and odd modes [66]. One can see from fig. 3 (b) that the transmission dips have redshifts as the outer radius increases where similar result for the filter of fig. 2 (a) was obtained for various outer radius [66].

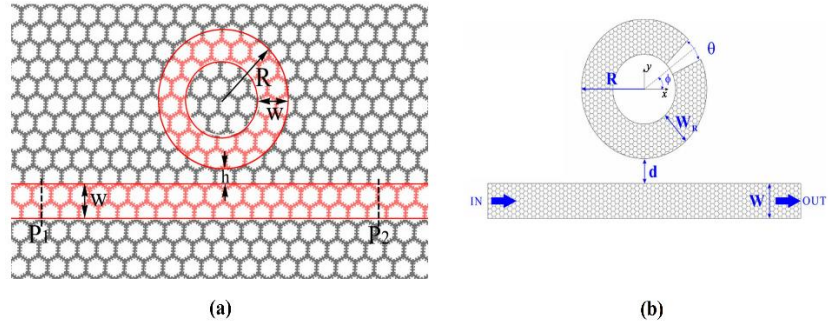

**Fig. 2**. (a) Band-stop filter constructed by a graphene ring resonator and a strip [65] (Reproduced by permission of AIP Publishing), (b) Tunable filter composed by a graphene strip waveguide and a graphene split ring resonator with split width corresponding to a sector angle $\theta$ [66] (Reprinted by permission from Springer Nature)

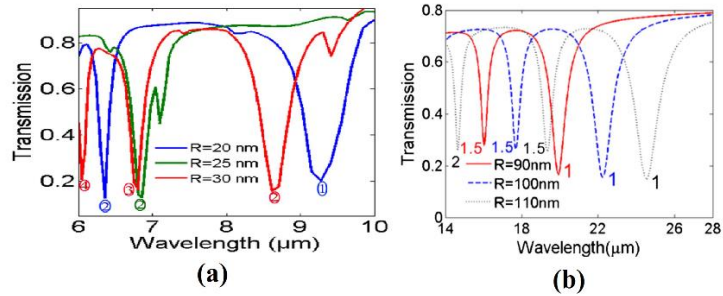

**Fig. 3**. (a) Transmission spectrum as a function of wavelength for the various radius (*R=20,25,30 nm* and the width of the strip is supposed *w= 10 nm* [65] (Reproduced by permission of AIP Publishing), (b) Transmission spectrum as a function of wavelength for the various radius (*R=90,100,110 nm*); other geometrical parameters are supposed *w=$w_R$= 50 nm, d= 20 nm,*
$\varphi = 90^0, \theta = 10^0$ [66] (Reproduced by permission from Springer Nature)

In [67], a tunable bandpass plasmonic filter based on a graphene nanodisk has been considered. In this filter, monolayer graphene was located on $SiO_2$/Si substrate, where the distance between the ribbon with the length of L and width of *w* has the distance of D with graphene disk [67]. At first, the authors studied the modal properties of the graphene ribbon as a function of ribbon width (d) at various THz frequencies and then proposed the filter with a graphene width of 20 nm to support single edge modes [67]. The transmission spectrum as a function of frequency and $E_z$ field distributions at different transmission peaks have been plotted in the article and concluded that the input power could be coupled to nanodisk at peak frequencies [67]. In the rest of the paper, a novel plasmonic filter based on a flat-disk resonator has been introduced and discussed, where the structure is composed of a round disk with a radius of *R* being elongated in the length of *h* [67].

*3.2 Planar Filters based on Graphene Nano-Ribbon (GNR)*

Graphene Nano-Ribbon (GNR) is a graphene sheet with a finite width in the nano dimension. In this sub-section, we will study the planar filters based on GNRs. The first planar filter was presented by M. Danaeifar et al. in 2013 [68]. This device was composed of the graphene sheet placed on the dielectric substrate with edge voltage [68]. Two ports at both sides of the structure had been defined to simulate the structure and measure the S-parameter and transmission coefficients [68]. Indeed, this structure is the simplest planar filter reported in the literature. The authors emphasized that the central frequency and quality factor of the filter could be tuned by changing substrate thickness, the gate voltage, and temperature [68].
Hong-Ju Li and his coworkers have been introduced two planar filters based on GNRs in 2014 [69] and 2015 [70], respectively. The schematic of both filters is illustrated in fig. 5 [69, 70]. In the first structure, an ultra-narrow band filter by two coplanar graphene strips located on the dielectric with refractive index *n* has been proposed while in the second filter, plasmonic band stop filter with a nanoribbon waveguide is placed in a distance of short graphene strip and both of them are deposited on the substrate with refractive index *n* [69, 70]. The authors used FDTD to numerically investigate both structures and both filters work based on edge modes of graphene [69, 70]. A parametric study for considering the effect of various parameters on the transmission spectrum has been done in both articles [69, 70]. The authors expressed that these filters can be utilized in designing novel structures such as sensors [69, 70].

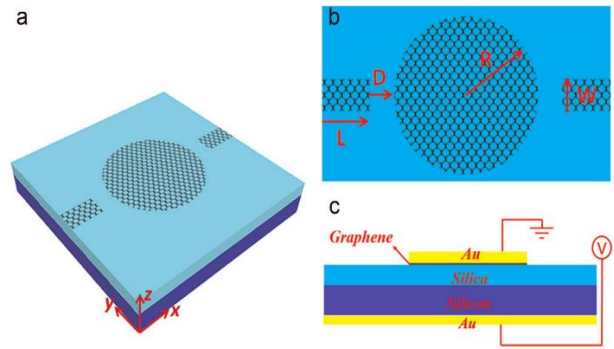

**Fig. 4**. (a) 3D view, (b) Top view and (c) cross-section of the tunable graphene filter [67].
Reproduced by permission from Elsevier.

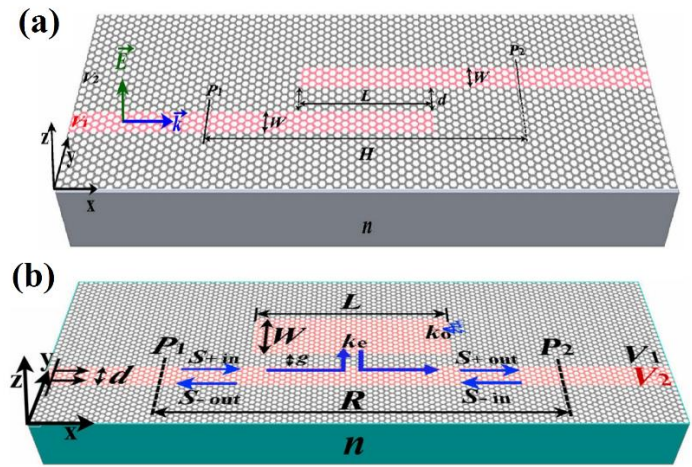

**Fig. 5**. (a) Ultra-narrow band filter by two coplanar graphene strips [69], (b) Plasmonic band stop filter with a nanoribbon waveguide and short strip [70]. Both figures are reproduced by permission from Springer Nature.

The article [71] focuses on designing a tunable graphene-based filter by using leapfrog alternating direction implicit finite difference time domain (leapfrog- ADI-FDTD) method and the details have been investigated in the context of the paper. This article is one of the designed plasmonic filters where the anisotropic graphene sheet with tensor conductivity (see relations (8) and (9)) [71].

Designing of plasmonic bandpass filters based on GNRs has been discussed in [72]. In this research, the authors first presented a graphene-based plasmonic filter in which two graphene nanoscale waveguides are coupled to a graphene ribbon [72]. The transmission and reflection spectrum of this filter are studied more precisely [72]. This work is one of the research articles that reports applied structures for nanocircuits in the mid-infrared region [72]. For instance, the authors introduced two applied structures based on the first filter: 1- graphene-based filter with the function of band selection 2- graphene-based filter with the function of $1\times 2$ power splitter [72]. The details of designing these structures are explained in [72].

Haidong Deng, et al. have proposed a bandpass filter based on coupled resonators, where the filter consisted of two coupled resonators and graphene waveguides [73]. The authors have used an intermediate graphene sheet between two resonators to achieve a high order filtering [73]. They have reported that the central wavelength of the passband filter can be tuned from $9-10\,\mu m$ by varying the chemical potential of graphene [73].

The article [74] introduces a graphene-based racetrack-resonator that works as an add/drop filter. This filter is composed of two directional couplers, graphene waveguide and racetrack resonator with $90^0$ bends [74]. The results of COMSOL simulations for the designed filter indicate that the filter can be a good candidate device in future integrated photonics in the infrared region [74]. Jie Huang, et al proposed a tunable flat-top bandpass filter based on graphene in [75]. In this filter, the authors have applied six graphene layers to enhance the effective mode index ($n_{eff}$) [75]. The achieved extinction ratio for drop port is larger than 40 dB and the insertion loss is less than 1.5 dB [75]. Another famous technique for designing a plasmonic bandpass filter is based on Fabry-Perot GNR, which has been studied in [76]. In this work, the filter consisted of two GNR waveguides that coupled to three perpendicular GNRs. These three perpendicular GNRs composed a Fabry-Perot resonator and all of these nanoribbons are placed on $SiO_2/Si$ substrate [76]. A parametric study has been done in the rest of the paper to determine the influence of chemical potential and geometrical parameters on the transmission spectrum [76].

*3.3 Periodic Filters*

In this sub-section, we will review the articles on periodic graphene filters. Diego Correas-Serrano et al. [54] proposed a graphene-based tunable low-pass THz filter where a graphene strip with width *w* is located on a dielectric and a set of polysilicon ac gating pads [77]. Transverse magnetic (TM) plasmon modes are propagated in the structure and different dc bias voltages to gating pads are applied to control the guiding characteristics [77]. Therefore, the structure was composed of cascaded transmission lines [77]. The authors reported the simulation and numerical results which indicated that the cut-off frequency can be tuned in the THz band [77]. It has been concluded that the designed reconfigurable filter can be utilized in THz plasmonic circuits such as sensors [77]. A tunable notch filter by using Kerr nonlinearity in graphene nanoribbon array has been studied in [78]. In this work, the structure was illuminated by incident light and the modulation of transmission based on the nonlinear Kerr effect of graphene was investigated more precisely [78].

Victor Dmitriev and his coworkers presented two periodic THz filters based on graphene. The first research structure was composed of the periodic square array, where two coaxial graphene rings have been located at the opposite sides of the dielectric and the whole structure was illuminated by an incident wave [79]. Each ring had various dimensions and therefore different resonant frequencies. As discussed in this work, the radiated waves from two rings are a destructive effect and high transmission occurs which is called Fano resonance [79]. The simulation results of COMSOL showed the quality factor of *Q=5* at *f=0.8 THz* for $\mu_c = 0.6\,eV$ [79]. The authors reported a parametric study to consider the effects of a dielectric substrate and chemical potential on the transmission and reflection spectrum of the filter in the rest of the paper [79]. The second article presented by Victor Dmitriev et al. has been considered a periodic FSS THz filter based on graphene [80]. Two graphene crosses are placed at both sides of the dielectric instead of two rings [80]. Again, the incident wave illuminated the structure and current distributions had opposite directions [80], similar to [79].

Periodic graphene filters can be designed by the THz metamaterials theory. For instance, K.Yang et al. designed the periodic graphene filter based on metamaterial where the unit cell was a metallic array with gap and graphene strip were located in gaps [81]. The simulated results indicated that the resonance frequency of the filter could be changed from 0.12 THz to 0.2 THz by varying the conductivity of graphene [81]. Another article about the metamaterial graphene filter is studied by the FDTD method in [82]. The unit cell of the proposed filter had two mutual parallel graphene ribbons placed on the center of a quartz substrate and TM SPP waves were existed by a normal incident wave [82]. The authors are considered the parallel asymmetric graphene ribbons in the end part of the article and then discuss applications of asymmetric GR filter as a sensor or switch [82].

Hong-Ju Li and his coworkers have been published two research articles about periodic graphene filter, as exhibited in fig. 6. The first structure consists of two graphene sheets on both sides of the periodic Dielectric-Air layers [83]. The applied method for study the structure is FDTD in both periodic filters [83, 84]. The second structure is constructed by a graphene sheet located on the periodic air trenches in the silicon waveguide [84]. In both filters, the periodic section acts as Plasmonic Bragg Reflector and the transmission spectrum of structures are depicted to consider the filtering effect [83, 84].

Bin Shi et al. are focused on band-stop filters based on periodically modulated graphene [85]. In the designed filter by them, the $SiO_2$ covers the periodic Si grating with a periodicity of $D$ and the graphene layer has been deposited on the $SiO_2$ layer [85]. The simulation results demonstrated a filter with good characteristics such as modulation depth of -26 dB [85]. A switchable microwave filter based on graphene has been introduced in [86]. The filter was composed of the graphene layer on a $SiO_2$/Si substrate and a periodic slot array in a metal plate has been placed between a dielectric and $SiO_2$ layer [86]. The structure was simulated in CST and transmission and reflection diagrams confirmed that the bandpass resonance could be changed by varying chemical potential of graphene and therefore the structure was a tunable filter [86].

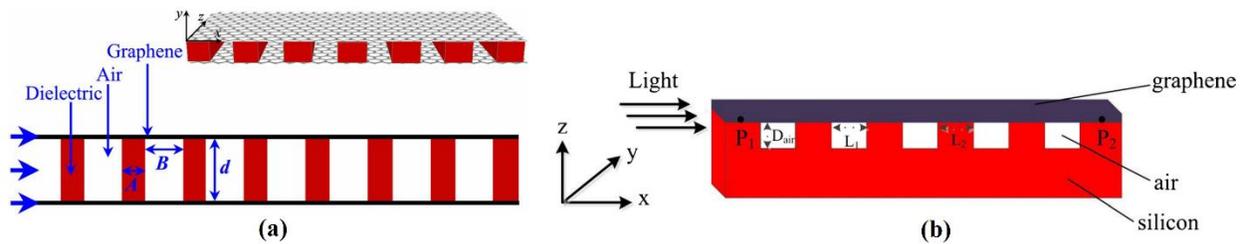

**Fig. 6**. Two periodic graphene filters introduced by the Ling-Ling Wang group: (a) two graphene sheets are located at both sides of a periodic Dielectric- Air layers [83], (b) graphene plasmonic Bragg reflector filter [84]. Both figures are reproduced by permission from Springer Nature.

## 3.4 Other Types of Filters

In this sub-section, we are going to introduce plasmonic graphene filters that do not put in any of the mentioned categories. One of them is THz angular/frequency filters designed by the modified Kretschmann–Raether configuration [87]. The graphene layer has been located in the middle of the structure and the whole filter is excited by an incident plane wave [87]. By exciting the transverse magnetic plasmons on graphene, the transmission resonances occur at THz regime where resonant wavelength and angle can be tuned by $\mu_c$ [87]. The authors have applied the modified transfer matrix method to calculate reflection and transmission coefficients [87]. The authors have been demonstrated that there is one dip in the transmittance spectrum in the angle-dependent reflectance in absence of graphene sheet while by exciting TM SPPs in the inserted graphene layer, two dips appear [87]. The second dip is deep and narrow and it exists due to TM SPPs [87]. The reader can refer to [87] for more consideration.

Article [88] presented a tunable bandpass filter based on the graphene-metal hybridization. Two metal trapped lines and five metal-graphene hairpin resonators constructed the filter [88]. The simulation results demonstrated that central bandpass frequency can be controlled in 2.96-3.12 THz [88]. Researchers are interested to utilize graphene properties such as flexibility and transparency to design novel devices in GHz frequencies. For instance, Jinchen Wang and his co-workers have introduced a microwave filter based on graphene at 5.8 GHz [89]. The authors fabricated the filter and measured results obtain insertion loss of 1.6 dB and return loss of 33.4 dB [89]. Also, the filter is tunable and the bandwidth and central frequency of this filter can be tuned by chemical potential [89].

One of the interesting research areas in plasmonic graphene filters is designing filters based on photonic crystals (PhC). In [90], one-dimensional graphene PhC filters were proposed by Yizhe Li et al., where single-layer defect and the dual-layer defect has been used in one-dimensional graphene PhC [90]. A TM polarized incident wave illuminated the graphene PhC filters [90]. The authors have started to analyze the structure by transfer matrix method where the details are in [90]. The transmission spectrum was plotted in the article at normal incidence and figures indicated that inserting single and two defect layers could appear one and two transmission peaks in each bandgap, respectively [90]. At the end of this article, the authors discuss applications of this kind of filter in THz narrow multi-band systems [90].

## 4. Perspectives and Challenges

Since graphene filters are designed based on the optical conductivity of graphene, they provide some useful benefits, which are important in mid-infrared frequencies. The most popular advantage of graphene-based plasmonic filters is their tunability. As mentioned before, the conductivity depends on various parameters such as electric bias, chemical doping, and magnetic bias. Therefore, tunable graphene-based filters can be designed for various applications. Another interesting feature of graphene-filters is their good confinement in the mid-infrared region, which can be utilized in designing compact nano-scale devices.

However, graphene filters suffer from the large propagation loss, due to the lossy feature of the graphene layer. This disadvantage makes the implementation of filters impractical. To overcome this, graphene technology can be integrated with other high-index contrast filters such as the SOI platform. It must be emphasized that the fabrication of graphene-based structures has standard processes in commercial and many chemists work on the various methods of graphene fabrication. Graphene filters have many useful applications in mid-infrared and THz technologies. They can be utilized in the optical switch, plasmonic waveguide, PhC structures, and unidirectional devices.

## 5. Conclusion

In this paper, a historical review of plasmonic graphene filters was presented. Due to the importance of graphene conductivity in designing plasmonic filters, the conductivity of graphene in presence of electrostatic or magnetostatic bias was considered at first. Then, various types of graphene-based filters were introduced and their principal operations were studied more precisely. In most of these filters, the graphene was biased electrically. Graphene filters exhibit some useful features such as tunability and good confinement in the mid-infrared region. They have many applications in the THz region such as a plasmonic waveguide and can be utilized in PhC structures and unidirectional devices.

[88] Y. Yao, X. Cheng, S.-W. Qu, J. Yu, and X. Chen, "Graphene-metal based tunable band-pass filters in the terahertz band," *IET Microwaves, Antennas & Propagation,* vol. 10, pp. 1570-1575, 2016.

[89] J. Wang, Y. Guan, and S. He, "Transparent 5.8 GHz filter based on graphene," in *Microwave Symposium (IMS), 2017 IEEE MTT-S International*, 2017, pp. 1653-1655.

[90] Y. Li, L. Qi, J. Yu, Z. Chen, Y. Yao, and X. Liu, "One-dimensional multiband terahertz graphene photonic crystal filters," *Optical Materials Express,* vol. 7, pp. 1228-1239, 2017.